%% file: TS_JINST.tex
\pdfoutput =1
\documentclass{JINST}
\let\ifpdf\relax
\usepackage{siunitx}
\usepackage{color,amsmath}
\usepackage{booktabs}
\usepackage[normalem]{ulem}
\usepackage{lineno}
\usepackage[style=base]{caption}
\usepackage[style=base]{subcaption}
\title{A 260-Liter Test Stand for Liquid Argon R\&D}
\author{Yichen Li$^a$\thanks{Corresponding author.},
Aleksey Bolotnikov$^b$,  Milind Diwan$^a$,  Jay Hyun Jo$^a$, Steven Kettell$^a$, Steven Linden$^b$, Xin Qian$^a$, Matteo Vicenzi$^a$ and Chao Zhang$^a$ \\
\llap{$^a$}Physics Department,\\
  Brookhaven National Laboratory, Upton, NY 11973, USA\\
\llap{$^b$}Instrumentation Department,\\
  Brookhaven National Laboratory, Upton, NY 11973, USA\\
E-mail: \email{yichen@bnl.gov}}
\abstract{We describe the design and  performance of a 260-liter liquid argon (LAr) cryogenic test stand  for liquid argon detector research and development at BNL. The system uses gas-phase argon purification with continuous pump-free circulation, in which boil-off argon gas is purified, recondensed, and returned to the cryostat by gravity without a mechanical recirculation pump; it also incorporates an upgraded condenser that increases the effective thermal contact area by a factor of 13 relative to the previously developed 20-liter system reported perviously. A liquid argon purity monitor is installed to measure the electron lifetime directly in LAr, enabling quantitative characterization of charge attenuation due to electronegative impurities. Under the operating conditions reported here, the
demonstrated electron lifetime is 0.5 ms. The system is designed to enable rapid iteration of detector components in complete operational cycles, including pump-down, leak verification, cryogenic fill, stable operation, and warm-up, which can be completed within 7 days. Such a fast turnaround time, together with the medium-scale liquid volume and direct purity diagnostics, makes the facility well suited for testing and refining detector designs in support of large liquid argon time projection chamber (LArTPC) experiments.}
\begin{document}
%\linenumbers
\input{intro.tex}

\section{System Description}\label{sec:description}
\input{cryo.tex}

\input{operation.tex}

\input{purity.tex}
\input{comparison.tex}

\input{summary.tex}\label{sec:future}

\acknowledgments
This work is supported by Laboratory Directed Research and Development (LDRD)
of Brookhaven National Laboratory and U.S. Department of Energy, Office of Science and Office of High Energy
Physics under contract number DE-SC0012704.

\bibliographystyle{unsrt}
\bibliography{TS_JINST}{}
\end{document}
\end{document}

%% file: intro.tex
\section{Introduction}

Liquid argon detectors \cite{WILLIS1974221} achieve sub-ppb level purity using liquid recirculation systems with cryogenic pumps and high-capacity filters such in SBND, ICRAUS, and DUNE experiments~\cite{Antonello:2015lea,Adams:2013qkq,Amerio:2004ze} . While this approach has been successfully demonstrated at multi-ton and kiloton scales, it entails substantial infrastructure, including external purification units, continuous liquid pumping, and complex cryogenic services. Such systems are well suited for long-term detector operation; however, they are not optimized for rapid prototyping cycles in research and development (R\&D) environments because of their high cost, the complexity of the associated cryogenic plumbing and pumping systems, and excessively large scale, which results in cooldown and warm-up cycles that can extend from weeks to months. At the opposite extreme, very small laboratory cryostats do not fully reproduce the thermal behavior, impurity dynamics, and mechanical integration challenges encountered in larger detector volumes. An intermediate-scale liquid argon platform that maintains realistic cryogenic and purity conditions while minimizing infrastructure complexity is therefore highly desirable for detector component development\cite{LAPD,MTS,argontube}.

A previously developed 20-liter test stand  at BNL demonstrated that the required high-purity of sub-ppb level impurity concentration can be achieved using gas-phase purification alone, without liquid recirculation pumps, by continuously purifying argon boil-off gas and returning condensed purified liquid to the main dewar~\cite{Li_2016}. That system allowed for pump-free circulation of small volumes, but purity characterization relied primarily on gas analyzers and indirect lifetime estimates, which limits its applicability for detector prototyping.

To investigate a system with a larger applicable volume, we designed and constructed a 260-L system to demonstrate that gas-phase purification alone can sustain millisecond-scale electron lifetimes in a larger liquid argon volume suitable for medium-scale detector R\&D, while preserving operational simplicity and rapid turnaround.

The 260-L system relies on the pump-free circulation scheme while incorporating three key developments required for scaling. First, the condenser is redesigned as a multi-tube heat exchanger that increases the effective thermal exchange area by a factor of thirteen relative to the earlier 20-liter configuration, improving condensation efficiency and pressure stability. Second, a liquid argon purity monitor is installed to measure the electron lifetime directly in the liquid, providing a quantitative and application-relevant purity metric. Third, the system is explicitly engineered for rapid operational cycling; a complete cryogenic sequence, including vacuum pump-down, leak verification, cryogenic filling, stable operation, and warm-up, can be completed within 7 days.

Together, these features establish a medium-scale liquid argon platform that combines realistic detector conditions with laboratory-kind flexibility. The system enables iterative development of high-voltage feedthroughs, field cage structures, charge readout configurations, photon detection systems, and material compatibility studies in support of large LArTPC experiments. Throughout this paper, purity is quantified primarily by the charge attenuation as it drifts through the liquid argon volume. This is encapsulated in the electron lifetime constant $\tau$ (see Sec. \ref{sec:electron_lifetiem}) For meter-scale drift distances, a sub-ppb level impurities concentration level is required. 

This paper is organized as follows. Section 2 describes the cryogenic system and its components.  Section 3 reports system operation and performance. Section 4 presents the purity monitor and electron lifetime measurements. Section 5 compares the 260-L system with the earlier 20-L system, and Section 6 provides a summary and discussion.

%% file: cryo.tex
\subsection{Cryogenic System Overview and Operation}

\begin{figure}[!htbp]
    \centering
    
    \begin{subfigure}[b]{0.72\textwidth}
        \centering
        \includegraphics[width=1.05\textwidth]{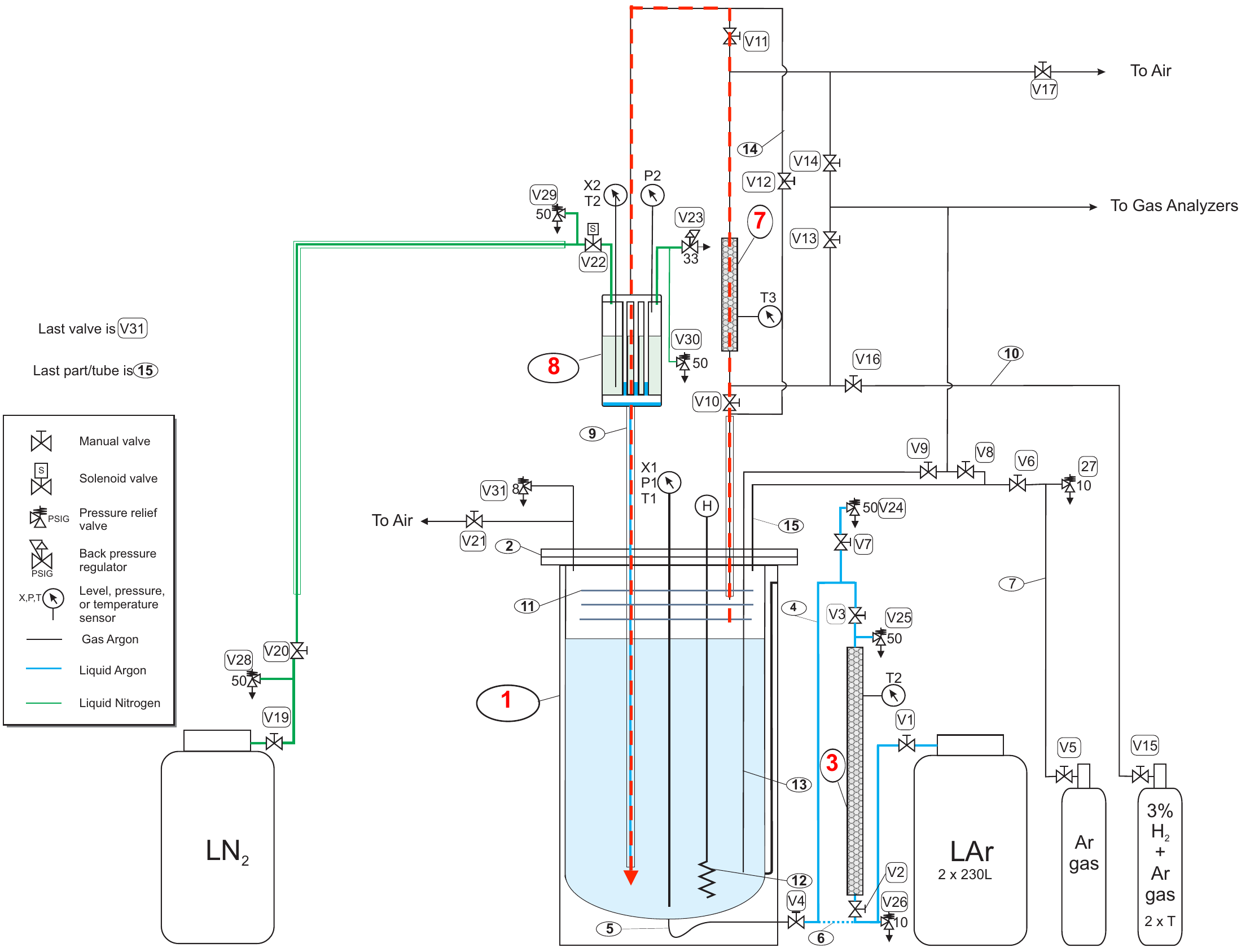}
        \caption{}
        \label{fig:PID}
    \end{subfigure}
    \hfill
    \begin{subfigure}[b]{0.25\textwidth}
        \centering
        \includegraphics[width=1\textwidth]{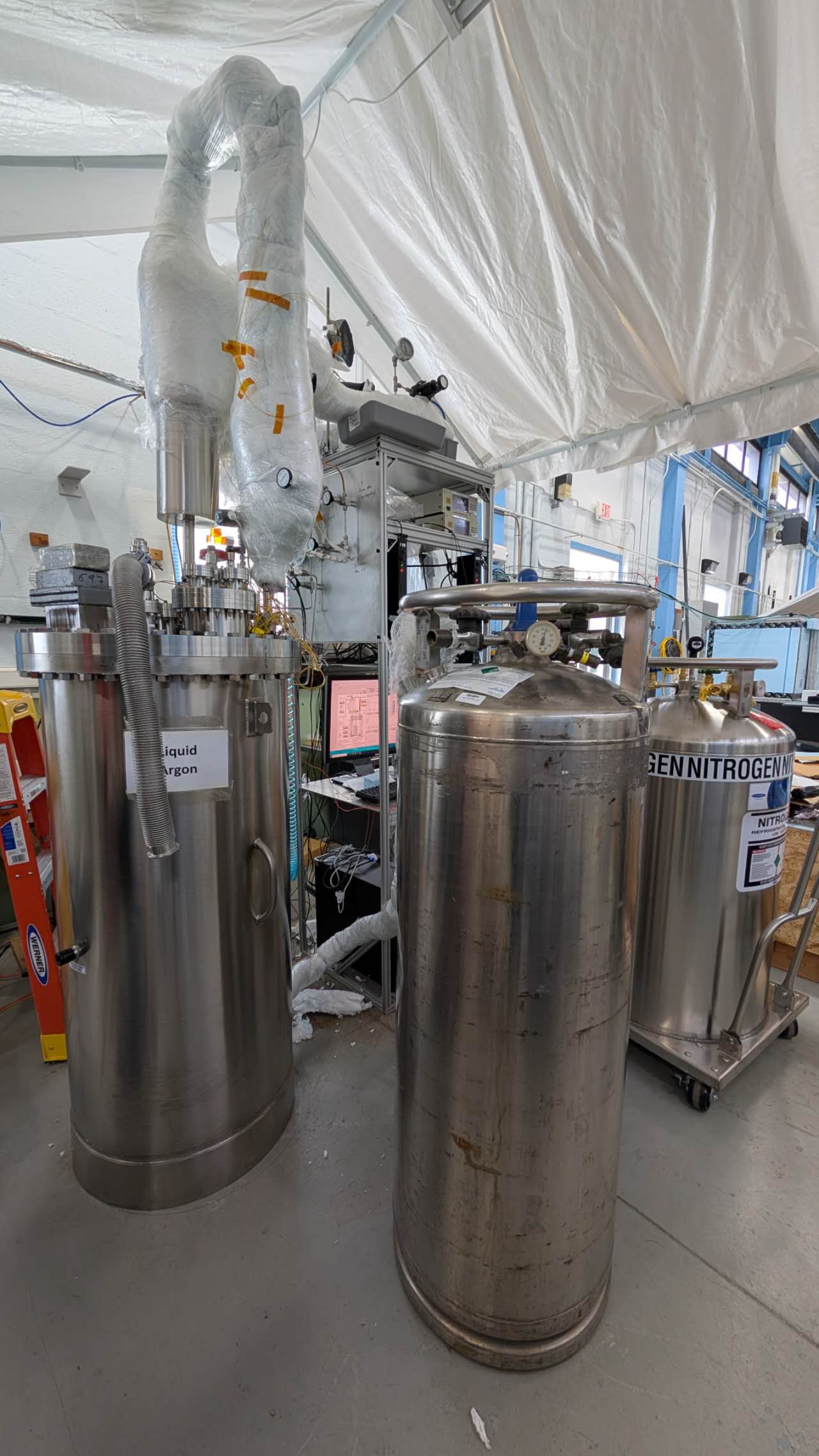}
        \caption{}
        \label{fig:206lphoto}
    \end{subfigure}
    
    \caption{(a) Schematic of the 260-L system. The essential components are highlighted in red: main cryostat (1), purifier (7), condenser (8), and inline filter (3). The pump-free circulation path is shown by the red dashed line. (b) Photograph of the setup, with the main dewar on the left, the LAr supply dewar in the middle, and the LN2 supply dewar on the right.}
    \label{fig:schematic}
\end{figure}

The cryogenic system is built around a vertical Multi-Layer Insulation (MLI) cryostat with a nominal capacity of 260 L. Fig.~\ref{fig:schematic} shows a schematic representation of the system.  The inner
vessel is 22 inches in diameter and is sealed by a 27~inch top access WireSeal flange that supports multiple instrumentation and service feedthroughs. The cryostat has been pressure-tested to 30~psig and is administratively limited to a maximum allowable working pressure of 15~psig. The top flange accommodates feedthroughs for high-voltage, readout electronics, temperature and pressure sensors, liquid level instrumentation, and an optical feedthrough used for ultraviolet light delivery when operating electron sources or the purity monitor.

%\begin{figure}[htbp] 
%\centering
%\includegraphics[width=0.95\textwidth, angle=0]{./figures/PID_v6.png}
%\caption{Schematic of the 260-L system, including the essential components marked in red:
%7) main cryostat, 2) purifier, 8) condenser and 3) inline filter, with the
%pump-free circulation path with shown in the red dashed line.}
%\label{fig:schematic}
%\end{figure}

This new 260-liter system adopts the same cryogenic concept as the previously developed 20-liter system~\cite{Li_2016}, with significant scale-up by one order of magnitude. Compared to the 20-liter system, to ensure high-purity performance, all seals in the system were implemented as metal-to-metal connections, using VCR or Swagelok fittings, thereby avoiding joints that require Teflon tape, O-rings, or vacuum grease. The cooling relies on balancing the heat load into the liquid argon with the cooling power provided by condensing boil-off argon gas in a liquid nitrogen LN$_2$-cooled heat exchanger. Heat leak into the cryostat vaporizes a fraction of the liquid argon, increasing the argon vapor pressure and driving a steady gas flow through the purifier cylinder and onward to the condenser. The condenser removes the latent heat of condensation from the argon gas using liquid nitrogen at a controlled saturation temperature, producing purified liquid argon that returns to the cryostat by gravity through insulated tubings. In steady state, the argon mass flow rate through the circulation loop is set by the system heat load and the effectiveness of the condenser, and the system circulates argon continuously without using a mechanical recirculation pump. This pump-free architecture removes a moving cryogenic component from the argon circulation and avoids the high cost, plumbing, controls, vibration, and maintenance associated with a cryogenic recirculation pump. This feature is especially useful for an R\&D stand that is opened frequently for detector
modifications.

The temperature of the LN$_2$ bath is controlled by regulating its pressure using a back-pressure regulator driven with compressed air. Because nitrogen and argon have different vapor pressure curves, LN$_2$ must operate at an elevated pressure in order to reach temperatures near the argon saturation region relevant for typical operation. A pressure-regulated LN$_2$ bath therefore provides an effective and tunable refrigeration source for condensing argon boil-off while maintaining stable liquid argon temperature and pressure.

\begin{figure}[htbp] 
\centering
\includegraphics[width=0.45\textwidth, angle=0]{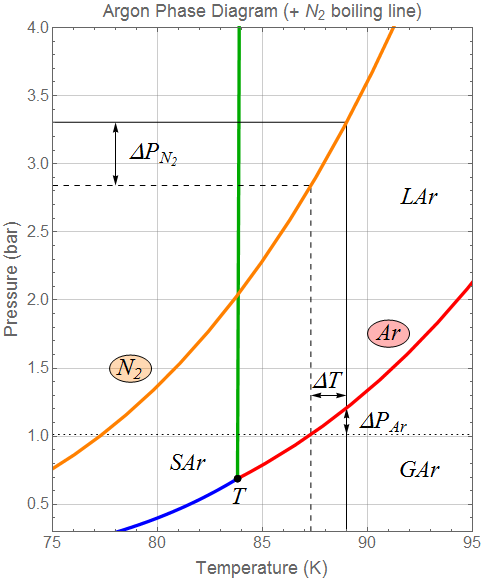}
\caption{Phase diagram of Argon together with the boiling line of N$_2$. 
The dashed black line shows the LAr at normal boiling point of 87.3 K. The 
solid black line indicates the normal operating temperature at 89 K.
$\Delta$P$_\text{Ar}$ and $\Delta$P$_{\text{N}_2}$ illustrates the
pressure difference associated with the temperature variation $\Delta$T. 
Due to different thermal properties, LN$_2$ is used as the source
of refrigeration of LAr in this system. Regulating the LN2 pressure sets the condenser
temperature and therefore the LAr operating point}
\label{fig:heat_ex}
\end{figure}

Fig.~\ref{fig:heat_ex} provides a phase-diagram view of the operating principle. This figure is included primarily to clarify the rationale for operating LN$_2$ at elevated pressure and to illustrate how LN$_2$ pressure control maps onto LAr temperature control. The liquid-argon (LAr) line in blue and the liquid-nitrogen (LN$_2$) line in orange on the left side of the figure represent saturated liquid-vapor conditions across the heat-exchange barrier. The small temperature difference, $\Delta$T, ensures net heat transfer from argon to nitrogen: argon releases latent heat during condensation, while the LN$_2$ side absorbs this energy as latent heat of vaporization during phase change. 

The turnaround time for a complete cryogenic operation cycle is approximately 7 days, including system warm-up and opening, closure with pump-down and leak checking (~1 day), pump-and-purge (2–3 days), LAr filling (~1 day), achieving target purity (1–2 days), and finally draining, LAr return to the supply dewar, and warm-up/reopening (~1 day after completion of physics measurements).

\subsection{Upgraded Condenser Design}

As a major upgrade to the previous 20-L system, the condenser is constructed as a parallel tube heat exchanger with an outer vessel that contains pressurized LN$_2$ and an internal argon flow path in thermal contact with the LN$_2$ bath. In the earlier 20-liter design, the argon gas was condensed in a single \SI{2}{inch} inner tube within a \SI{6}{inch} LN$_2$ vessel. In the 260-liter system, the condensation assembly is redesigned as a bundle of 50 parallel \SI{0.5}{inch} tubes, which increases the effective thermal contact area by a factor of 13 while also improving flow distribution and condensation efficiency. Additional MLI insulation is applied around the bottom and side surface of the condenser to reduce parasitic heat leak and to improve overall thermal efficiency.

\begin{figure}[!htbp]
    \centering
    
    \begin{subfigure}[b]{0.48\textwidth}
        \centering
        \includegraphics[width=0.33\textwidth]{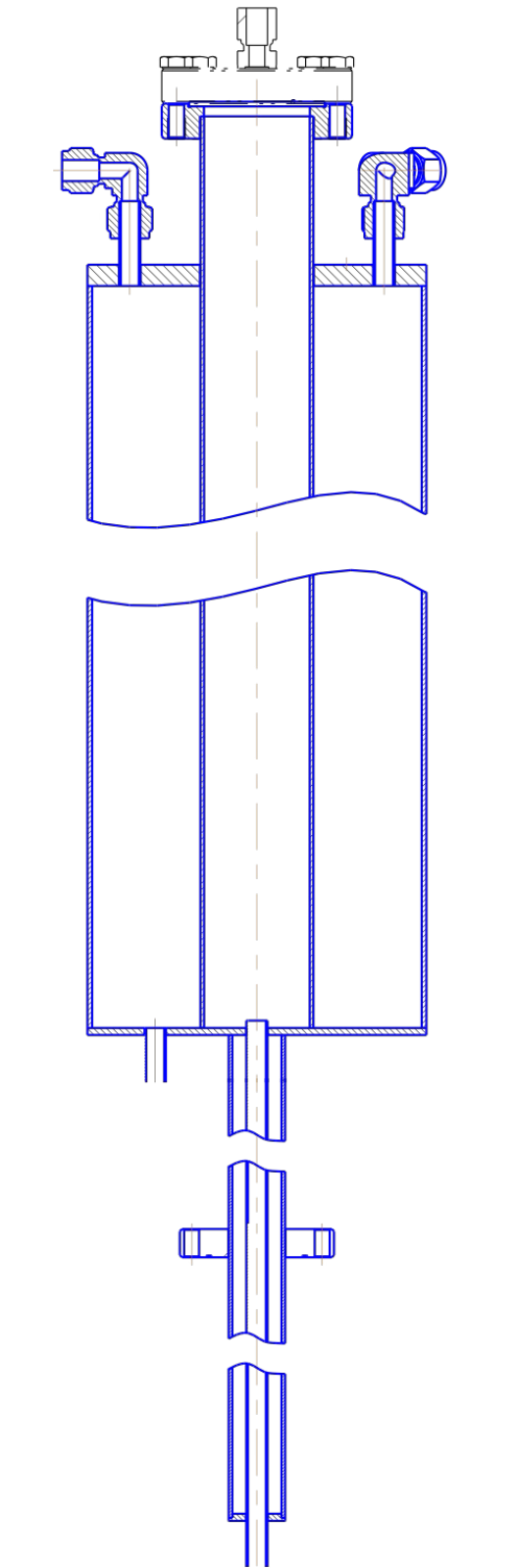}
        \caption{Structure 20-L system condenser with a single 2-inch inner tube within a 6-inch LN$_2$ vessel, no insulation on the outer jacket.}
        \label{fig:20l}
    \end{subfigure}
    \hfill
    \begin{subfigure}[b]{0.48\textwidth}
        \centering
        \includegraphics[width=0.25\textwidth]{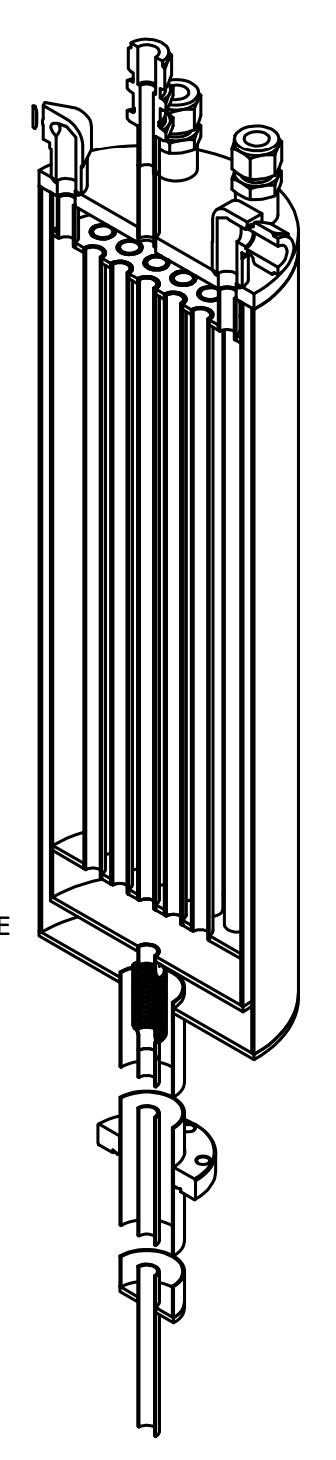}
        \caption{Structure of 260-L system condenser with 50 parallel 0.5-inch tubes, with MLI insulation on the bottom and side surface.}
        \label{fig:260l}
    \end{subfigure}
    
    \caption{Comparison between 20-L and 260-L condenser structures with the same height and inner diameter of 6-inch. The difference is at the size of inner heat exchange tubes.}
    \label{fig:condensers}
\end{figure}

The condenser vessel outer diameter is set to 6 inches. Although a larger diameter would improve performance, it would require compliance with pressure-vessel codes rather than pressure-piping standards, significantly increasing cost and complexity. The 6-inch diameter is therefore retained as a practical compromise. The comparison between the condensers in the 20-L and 260-L systems is shown in Figure \ref{fig:condensers}.

\subsection{Purification and Regeneration}

The purifier is a stainless steel cylinder filled with two materials that target the dominant electronegative contaminants in argon, namely water and oxygen. These impurities strongly absorb free electrons and degrade the charge signal over long drift distances more than other common contaminants in liquid argon. Water is removed using 13X molecular sieve via adsorption in its microporous zeolite structure. Oxygen is removed using a copper-based catalyst medium (GetterMax-233) via chemical reaction in which oxygen oxidizes metallic copper \cite{Gettermax}, according to
\begin{equation}
2Cu+O_2 \rightarrow 2CuO.
\end{equation}
These processes reduce impurity concentration in the passing argon gas, and the purified argon is returned to the cryostat as condensed liquid.

Since adsorption sites and reactive copper surface are consumed over time, the purifier must be regenerated periodically after 2 to 3 full cycle cryogenic runs. Regeneration is performed in situ by heating the purification vessel to  \SI{190}{\celsius}. During heating, high-purity argon is flowed to remove desorbed moisture from the molecular sieve. The copper medium is regenerated by flowing a reducing mixture of 98\% argon and 2\% hydrogen, which converts copper oxide back to metallic copper through
\begin{equation}
CuO+H_2 \rightarrow Cu+H_2O.
\end{equation}
The regeneration duration depends on impurity loading and flow configuration; in typical operation it requires approximately 24 hours to achieve the designated purity.

\subsection{Inline Filter}
An essential component for quickly achieving the required sub-parts-per-billion(ppb) purity condition for physics measurement is the inline filter (labeled as component 3 in the Fig.~\ref{fig:schematic}). Although commercial argon is specified as ultra-high purity, it typically still contains impurities at the parts-per-million (ppm) level. Because the circulation rate in our pump-free gas-circulation system is relatively low compared with that of liquid-circulation systems, the initial purity of the detector following filling is a dominant factor in determining the turnaround time. The inline filter is filled with the same filtering material used in the gas-circulation purifier. By conducting a single-pass filtration of the commercial argon during filling through the inline filter, the impurity concentration can be reduced immediately from the ppm level to the parts-per-billion (ppb) level, thereby enabling the system to reach sub-ppb purity within two days with continuous circulation.

\subsection{Plumbing and Instrumentation}

The circulation includes lines for argon boil-off extraction, purifier flow, condenser flow, and liquid return to the cryostat. Where practical, return lines are vacuum-jacketed to minimize heat leak and to avoid parasitic boiling and two-phase instabilities. External pipes above the cryostat flange are insulated using cryogenic insulation and protected with a moisture barrier to reduce ice formation and uncontrolled heat ingress.

The system is instrumented to monitor and control pressure, temperature, and liquid level in both the cryostat and the condenser. Liquid level is measured using a capacitance probe and is cross-checked using differential pressure measurements between the bottom of the dewar and the vapor region. The differential pressure method uses the relation $\Delta P = \rho g h$, so the measured pressure difference provides an absolute liquid height when the density is known from temperature and pressure. Temperature is measured using calibrated RTD probes, and pressure is measured using precision transducers. Calibration and verification are performed using fixed points such as the LN$_2$ boiling temperature and by cross-checking sensors under stable operating conditions.

%% file: operation.tex
\section{System Operation and Performance}
\subsection{Slow Control for Cryogenic Operation}

System operation is coordinated by a slow control and data acquisition framework implemented in LabVIEW. The control system continuously records key operational parameters, including cryostat pressure and temperature, condenser pressure and temperature, LN$_2$ level, and liquid argon level. The same framework also implements closed-loop logic for LN$_2$ batch filling and provides interlocks and alarms for off-normal conditions.

LN$_2$ filling is controlled using a cryogenic solenoid valve and a level gauge installed in the condenser. Rather than continuous transfer, which can increase nitrogen consumption due to heat leak into the transfer line, the system operates using a batch-fill strategy. In this mode, the condenser LN$_2$ level is maintained between a low threshold and a high threshold. When the level falls below the low threshold, the valve opens and the condenser is refilled until the high threshold is reached, at which point the valve closes. This strategy provides adequate thermal stability for measurements while reducing LN$_2$ consumption associated with continuously chilled transfer hardware. The batch-fill thresholds and the resulting fill-cycle time are system dependent. For the operating configuration used here, the low and high thresholds are 45\% and 80\% of the active gauge length respectively,
corresponding to a refill period of 1.5-2.0 hours.
\subsection{Operating Procedure and Rapid Turnaround}

A normal operational cycle begins with vacuum pump-down to remove residual gases and water vapor from internal volumes, followed by a helium leak verification to confirm leak-tightness at a sensitivity appropriate for high-purity operation. After leak verification, the purifier is activated or confirmed to be in an active state. The cryostat is then cooled down and filled with liquid argon while maintaining pressure below administrative limits. Once filled, the condenser is operated under pressure control and batch-filled with LN$_2$ to maintain stable thermal conditions. After the desired measurement campaign, the system is warmed back to room temperature to enable access for modifications and upgrades.  In
current operation, the full cycle from pump-down to
warm-up fits within 7 days, which is substantially shorter than the turnaround times typical of large cryogenic facilities that may take a few weeks to achieve designated purity level and is therefore well suited for rapid iteration of detector prototypes.

\subsection{Thermal Stability and Batch-fill Mechanism}

\begin{figure}[t]
\centering
\includegraphics[width=0.75\textwidth, angle=0]{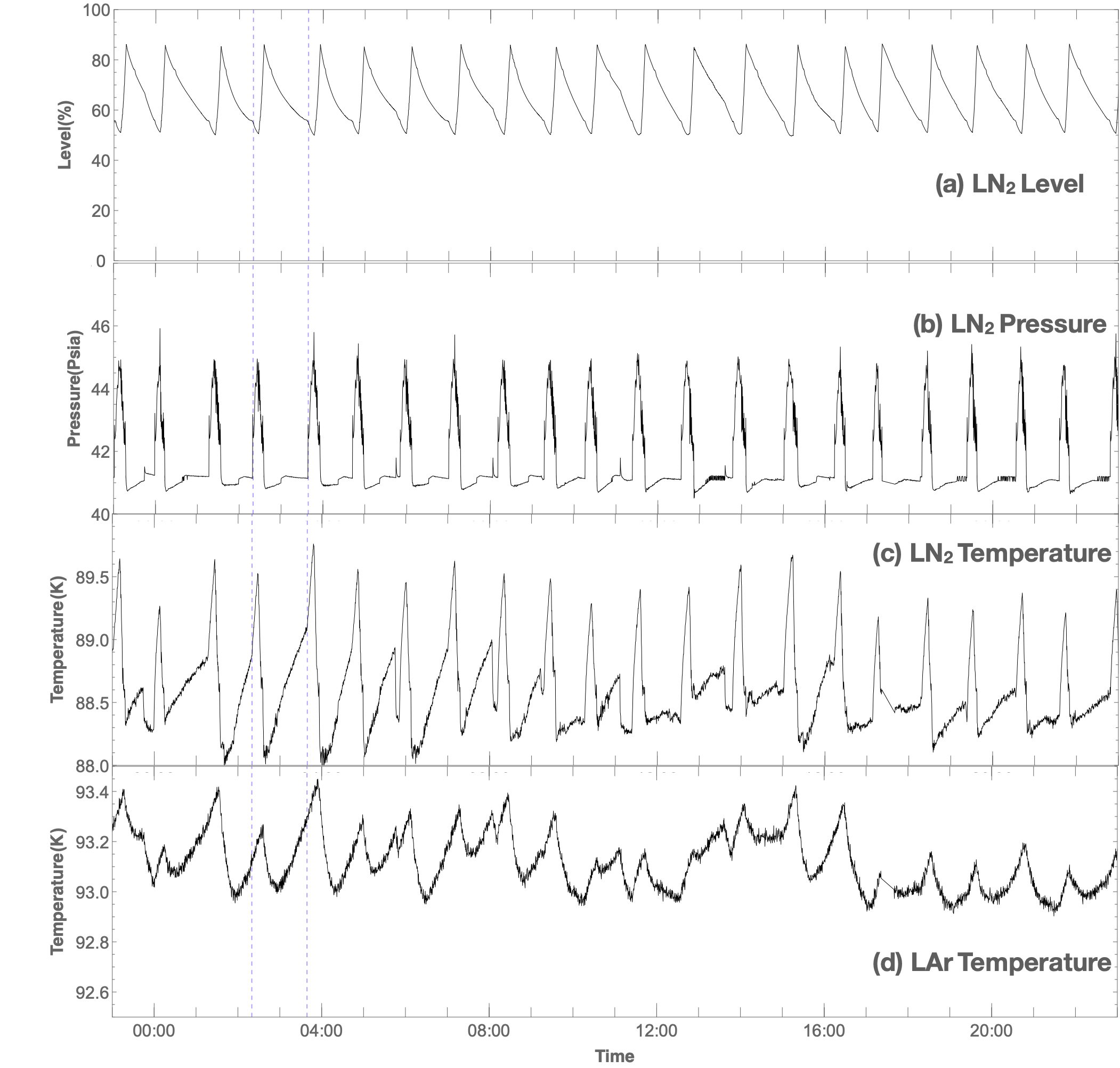}
\caption{Representative one-day history of condenser LN2
level, condenser pressure, LN2 temperature, and LAr
temperature during batch-fill operation: (a) liquid nitrogen level in the condenser
by volume, (b) liquid nitrogen pressure in the condenser, (c) liquid nitrogen temperature, and (d) LAr temperature are displayed during one day of the operation. The time range with the dashed line represents one
filling cycle. The two vertical blue dashed lines show the time interval of a filling cycle. See text for further discussions.}
\label{fig:stability}
\end{figure}

Thermal stability with temperature fluctuations below 1 K is primarily determined by the condenser operating point and the batch-fill mechanism. The system operating parameters over a 24-hour period, including the LN$_2$ level, pressure, and cryostat temperature, are shown in Fig.~\ref{fig:stability}, demonstrating the stability of the system.

A batch-fill mechanism is used to periodically fill the condenser to maintain the cryogenic operation. The filling is initiated when the LN$_2$ level in the condenser reaches a lower threshold and terminated at an upper threshold, with control provided by the LN$_2$ using feedback from a capacitance level gauge. The level thresholds are set at 55\% and 85\%, and a back-pressure regulator at 27 psig maintains the LN$_2$ temperature at 88.5 K. Pressure peaks correspond to the start of each filling cycle and arise from increased gas flow during filling. A brief undershoot in LN$_2$ level is observed due to transport delay, followed by a rapid rise during filling and a gradual decrease due to evaporation from argon condensation and thermal heat load. The LN$_2$ temperature slightly increases during filling, while the LAr temperature varies modestly during the cycle but remains stable within ~0.2 K. A full filling cycle lasts approximately 1.0 hour shown by the two vertical blue dashed lines in Fig.~\ref{fig:stability}.

During batch filling, transient changes in LN$_2$ pressure and level can introduce small perturbations to the condensation rate, which may manifest as small variations in cryostat temperature and pressure. The multi-tube condenser design increases heat transfer area and reduces sensitivity to such transients by maintaining effective condensation over a broader range of LN$_2$ levels.

A thorough understanding of the system's heat load is essential for proper 
cryogenic operation. The heat load is estimated using an approach similar to that used for the previous 20-L system, based on analytical calculations of the thermal properties of individual components. For the 260-L stand, the present estimate is 110 W during stable operation.

%% file: purity.tex
\section{Purity Performance}
\subsection{Purity Monitor System}

Because gas-phase purification directly processes the argon vapor above the liquid, and the primary source of impurity is the outgassing in the warmer surfaces, gas purification is particularly effective by removing impurities at the origin in gas volume. The approach can therefore maintain high purity in the liquid once the system has cooled and the dominant impurity sources have transitioned to the vapor region. In the 260-L system, a purity monitor based on the ICARUS design and ProtoDUNE-SP readout electronics is used to measure electron lifetime directly in LAr \cite{Amerio:2004ze,Abbaslu_2025}. 

The monitor follows the concept in which electrons are generated at a photocathode by ultraviolet illumination and drift through a defined distance under a uniform electric field before being collected at an anode. A grid is typically included to control the electrostatic boundary conditions and to separate induction effects from charge collection. The monitor provides two charge measurements: the charge associated with electron emission near the cathode, and the charge collected at the anode after drift. The ratio of these charges provides a direct measure of impurity-driven attenuation during drift.

\subsection{Electron Lifetime Measurement by Purity Monitor}\label{sec:electron_lifetiem}

The recorded purity monitor signal exhibits a prompt feature associated with the cathode signal, followed by a quiet drift interval, and then a delayed feature associated with charge arrival at the anode. The drift time $t$ is extracted from the measured separation between these features. When the drift distance is $d$ and the drift velocity is $v_d$, the expected relation is
\begin{equation}
t=\frac{d}{v_d}.
\end{equation}
In practice, the drift time is measured directly from the signal using a consistent timing convention, such as the time difference between the 50\% rising-edge points of the cathode and anode signals. This method reduces sensitivity to pulse height variations and electronic noise.

\begin{figure}[t]
\centering
\includegraphics[width=0.6\textwidth, angle=0]{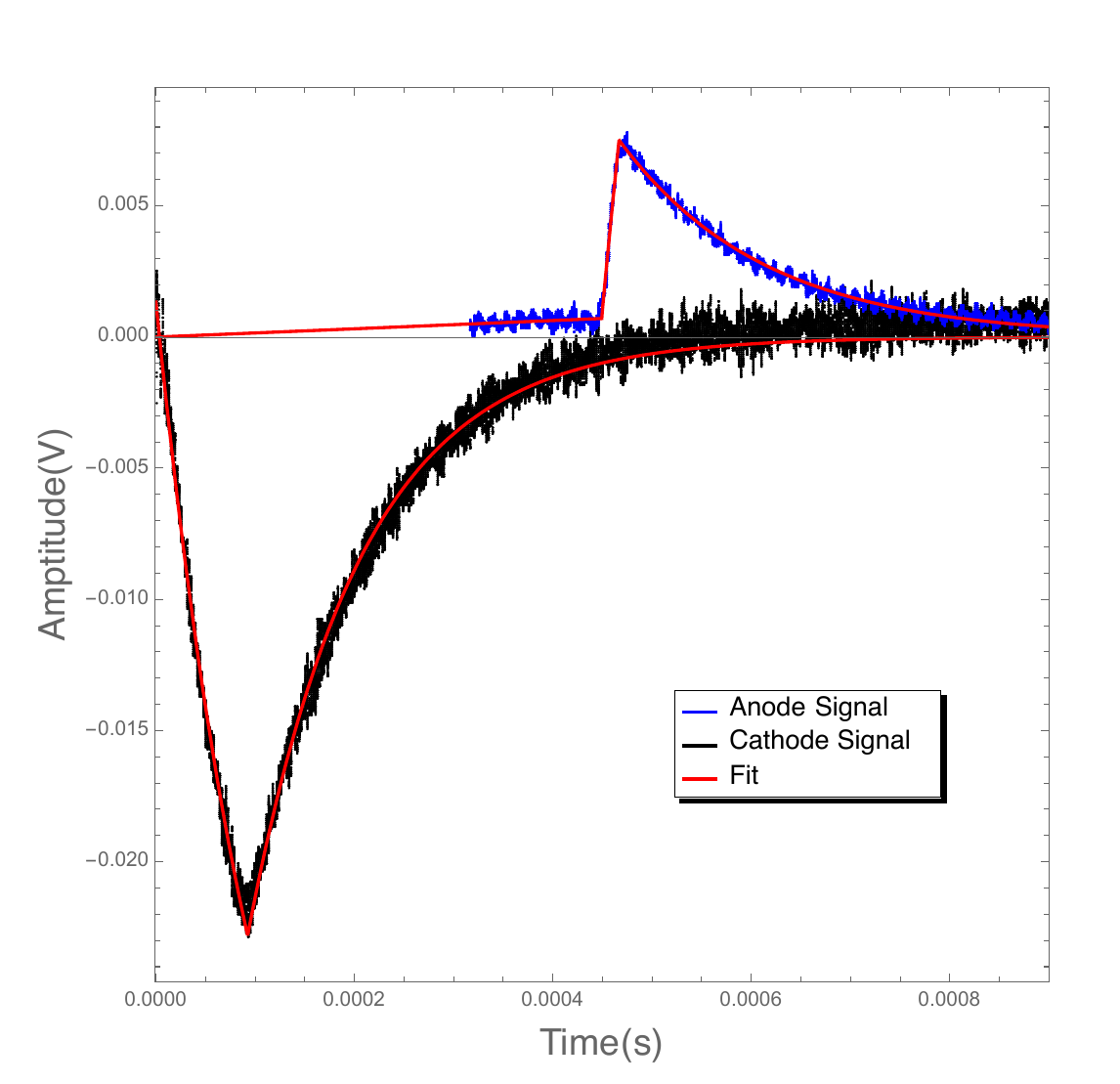}
\caption{Example purity-monitor waveform showing the
cathode signal, the drift interval, and the delayed anode
collection signal used to extract drift time and charge ratio.}
\label{fig:signal}
\end{figure}

Fig.~\ref{fig:signal} illustrates the characteristic signal structure from the purity monitor, including both the cathode and anode signals, from which the electron lifetime is determined using the same model as in ProtoDUNE-SP (See Sec 4.2 of \cite{Abbaslu_2025}).

The cathode-associated charge $Q_C$ and anode-collected charge $Q_A$, illustrated by the peak height of the signal shown in Fig. \ref{fig:signal} are extracted from the signal using baseline-subtracted pulse measurements. When a charge-sensitive preamplifier is used, the pulse height is proportional to the deposited charge through the preamplifier gain, which is calibrated using a known injected calibration charge through a reference capacitor. Calibration runs are performed periodically to verify gain stability and to quantify any temperature-dependent variations in the electronics response.

Electron attachment during drift produces exponential attenuation of the collected charge. 
The resulting negative ions drift very slowly ( $\sim$10$^5$
times slower than electrons) and the slow
signals they induce at the readout are outside the bandwidth of the detection electronics. The relation between cathode and anode charge measurements is therefore
\begin{equation}
Q_A/Q_c = e^{-t/\tau},
\end{equation}
and the lifetime can be determined by
\begin{equation}
\tau = \frac{t}{\ln(Q_A/Q_C)}.
\end{equation}
Signals are typically averaged over many ultraviolet illumination pulses to reduce electronic noise and to improve stability in the extracted $Q_C$ and $Q_A$ values. The analysis accounts for baseline offsets, and it applies consistent measurement windows or peak finding algorithms to avoid bias from pulse shape variations. Systematic effects arise from grid transparency, possible field non-uniformity within the drift volume, and gain drift in the readout electronics using the same analysis technique detailed described in \cite{Abbaslu_2025}. The best electron lifetime achieved is about 0.5 ms after 3 days of continuous circulation.

\subsection{Nitrogen concentration and photon detection considerations}

In addition to electronegative impurities with strong affinity to electrons such as oxygen and water, the concentration of nitrogen in liquid argon must also be considered, particularly for detector configurations that rely on scintillation light detection. While nitrogen has a negligible effect on electron lifetime at the ppm level compared to water and oxygen, it exhibits strong absorption of the vacuum ultraviolet (VUV) scintillation light emitted by liquid argon at 128~nm for $\mathcal{O}$(1 ppm)\cite{Acciarri_2010} concentrations. Elevated nitrogen concentrations can therefore reduce photon transmission and degrade the performance of light detection systems used for timing, triggering, and calorimetric measurements in LArTPC detectors.

Unlike oxygen and water, there is no cost-effective purification medium capable of selectively removing nitrogen from argon in cryogenic detector systems at present. Fortunately, the acceptable nitrogen concentration for photon detection applications is at the ppm level\cite{Acciarri_2010}, which is significantly less stringent than the sub-ppb requirements imposed on oxygen-equivalent impurities for charge collection.

Consequently, nitrogen control in the present system is achieved primarily through strict leak check procedures. Prior to filling with liquid argon, the cryostat and associated plumbing are evacuated and subjected to helium leak testing at a sensitivity of $10^{-8}$~standard~cc/sec. This level of leak tightness effectively suppresses atmospheric nitrogen ingress during operation.

\begin{figure}[t]
\centering
\includegraphics[width=0.67\textwidth, angle=0]{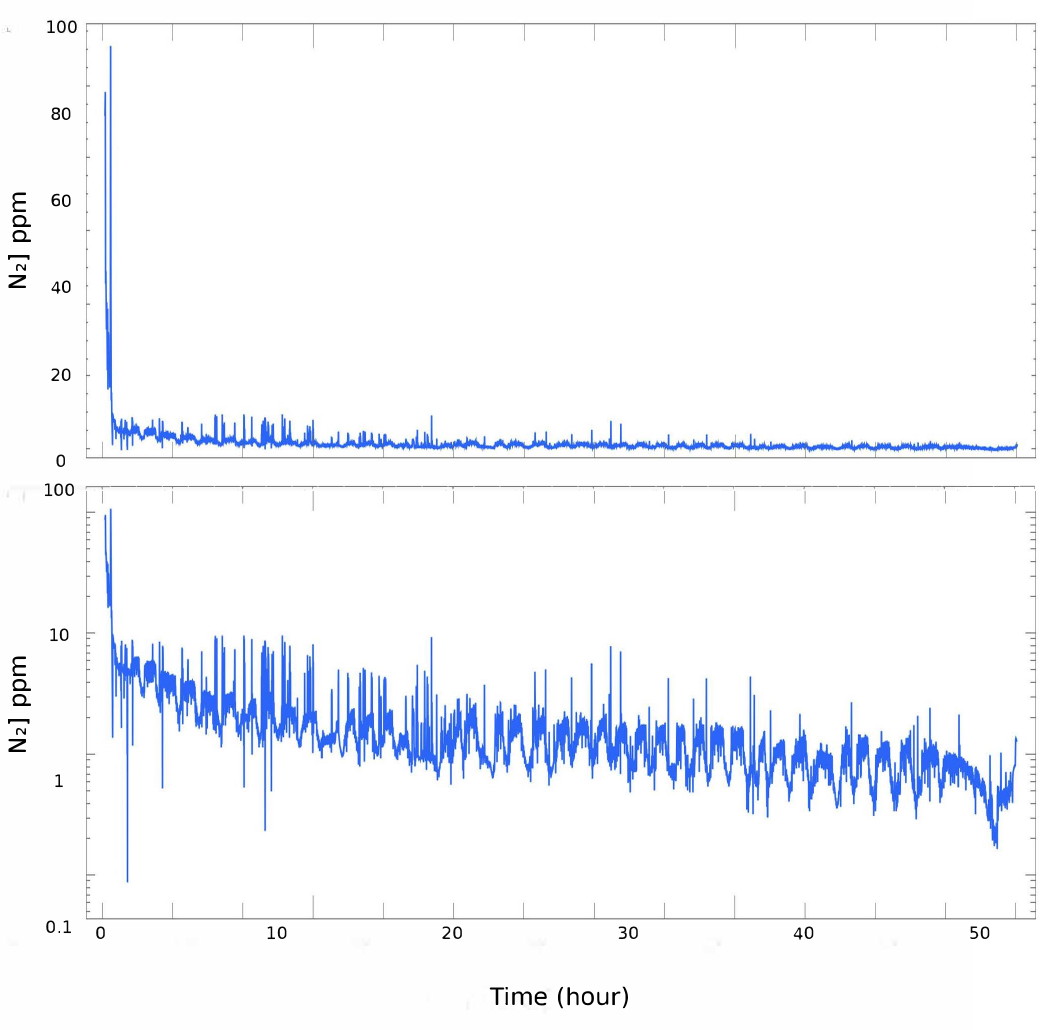}
\caption{ Nitrogen concentration as a function of operating time near the end of a cryogenic run. The upper panel presents the nitrogen concentration on a linear scale, whereas the lower panel shows the same data on a logarithmic scale for a clear illustration.}
\label{fig:n2con}
\end{figure}

Under nominal operating conditions with verified leak integrity, the nitrogen concentration in the 260-liter system remains around 1.0~ppm, as measured by gas analyzers sampling directly from the liquid shown in Fig.~\ref{fig:n2con}. This concentration is sufficiently low to permit photon detection studies less than 20\% VUV absorption \cite{Acciarri_2010}. The system therefore supports both charge-based and light-based detector research and development.

%% file: comparison.tex
\section{Comparison with the 20-L System}
In the 260-L system, the electron lifetime measured by the purity monitor increases over several days of continuous circulation and reaches 0.5 ms after 3 days (See Section \ref{sec:electron_lifetiem}). This 260-L system builds directly on the design philosophy of the earlier 20-L gas-purified test stand while addressing scale and diagnostic limitations. The increase in liquid volume enables more realistic detector prototyping, including instrumentation layouts and thermal behaviors that are not accessible in very small cryostats. The condenser redesign increases thermal exchange area by about a factor of thirteen, improving condensation effectiveness and reducing sensitivity to operating transients. The addition of a liquid argon purity monitor provides a direct measure of electron lifetime, which is the relevant purity parameter for charge-based detector operation. Finally, the 260-L platform is optimized for rapid cycling and frequent access at substantially larger volume.

\begin{table}[t]
\centering
\caption{Comparison between the earlier 20-L test stand and the 260-L system. }
\label{tab:comparison}
\resizebox{\textwidth}{!}{
    \begin{tabular}{lcc}
    \toprule
    Parameter & 20-L system\cite{Li_2016} & 260-L system \\
    \midrule
    Nominal LAr volume & \SI{20}{L} & \SI{260}{L} \\
    Purification approach & Gas-phase only & Gas-phase only \\
    Condenser geometry & \SI{2}{inch} inner tube in \SI{6}{inch} shell & 50$\times$\SI{0.5}{inch} tube bundle in \SI{6}{inch} shell \\
    Thermal contact area & 0.073 m$^2$ &  0.942 m$^2$\\
    Primary purity diagnostic & Gas analyzers (derived by concentration) & Purity monitor (direct $\tau$) \\
    Electron lifetime scale & $>\SI{0.4}{ms}$ (demonstrated) & $>\SI{0.5}{ms}$ (demonstrated)\\
    Turnaround time & $\sim$1 week & $\sim$1 week \\
    \bottomrule
    \end{tabular}
}
\end{table}

Compared to the 20-L system, the 260-L system is better suited
to detector-component R\&D because it provides a more
accessible volume, supports direct in-liquid purity
measurement, and retains a 7-day full-cycle turnaround.
Table~\ref{tab:comparison} summarizes representative differences.

%% file: summary.tex
\section{Discussion and Conclusion}

We have successfully constructed and operated a 260-liter liquid argon cryogenic test stand that employs gas-phase purification without liquid recirculation pumps. The upgraded multi-tube condenser design provides improved thermal stability and condensation efficiency, and the in-liquid purity monitor demonstrates an electron
lifetime of 0.5 ms under the operating conditions. These results demonstrate that ultra-high-purity liquid argon suitable for meter-scale drift distances can be achieved and maintained in a medium-scale system using gas-phase purification alone. An alternative approach by using a high-temperature rare-earth getter, is not cost-effective because such getters are consumable and cannot be regenerated. In addition, such a system requires active gas circulation with a pump, which further increases cost and may introduce an additional risk of contamination to the system. Compared to large pump-driven LAr systems, its strengths are low cost,
simpler plumbing, fewer active cryogenic components in the
argon loop, and faster access between runs. Its trade-off is
lower circulation throughput and a demonstrated lifetime of
0.5 ms rather than the several-millisecond values required for
long-drift detector operation, but sufficient for measurements at the drift scale of this system. Further improvements in purity performance can be achieved by increasing the size of the inline filter and the quantity of filtering material, thereby further improving the initial purity during the early stage of the purification process.

The ability to complete a full  cycle within one week makes this platform particularly well suited for rapid and iterative detector design and component testing. In addition to charge-based studies enabled by direct electron lifetime measurement, the controlled nitrogen concentration at the ppm level allows photon detection investigations without significant VUV light absorption effects. The 260-liter platform therefore supports both charge-readout and photon-detection studies and provides a flexible intermediate-scale facility to support technology development for future large LArTPC experiments.